\documentclass[sigplan,nonacm]{acmart}
\usepackage{amsmath,amsfonts}
\usepackage{algorithmic}
\usepackage{graphicx}
\usepackage{textcomp}
\usepackage{xcolor}

\usepackage[ruled,vlined]{algorithm2e} % or algorithmic, algorithmicx and other packages
\usepackage{algorithmic} % If you only use algorithm2e, you don’t need algorithmic
\newcommand{\microsubmissionnumber}{NaN}
\fancypagestyle{firstpage}{
  \fancyhf{}
  
  \fancyhead[C]{\vspace{10pt}\normalsize{MICRO 2024 Submission
      \textbf{\#\microsubmissionnumber} -- Confidential Draft -- Do NOT Distribute!!}\\\vspace{-25pt}} 
  \fancyfoot[C]{\thepage}
}

\usepackage[normalem]{ulem}
\usepackage{siunitx}            % typset units correctly
\usepackage{courier}            % standard fixed width font
\usepackage{comment}
\usepackage{helvet} % see www.ctan.org/get/macros/latex/required/psnfss/psnfss2e.pdf
\usepackage{xspace}
\usepackage{float}
\usepackage{amsmath}
\usepackage{xspace}
\usepackage{cleveref}
\usepackage{balance}
\usepackage{url}
\usepackage{siunitx}
\usepackage{comment}
\usepackage{enumitem}
\usepackage{listings}
\usepackage{subcaption}
\newcommand{\sys}{CXLMemSim\xspace}

\newcommand{\eg}{e.g.,\xspace}
\newcommand{\ie}{i.e.,\xspace}

  {\begin{enumerate}[itemsep=-0pt, parsep=0pt, topsep=0pt, leftmargin=2pc]}
  {\end{enumerate}}

  {\begin{list}{$\bullet$}%
    {\setlength{\parsep}{0pt}%
      \setlength{\topsep}{0pt}%
      \setlength{\leftmargin}{2pc}%
      \setlength{\itemsep}{1pt}}}
  {\end{list}}
\title{\sys: A pure software simulated CXL.mem for performance characterization}
  
\author{Yiwei Yang, Brian Zhao, Yusheng Zheng, Pooneh Safayenikoo, Tanvir Ahmed Khan, Andi Quinn}

\begin{document}
\begin{abstract}
\sys is a fast, lightweight simulation framework that enables performance characterization of memory systems based on Compute Express Link (CXL) .mem technology. CXL.mem allows disaggregation and pooling of memory to mitigate memory stranding (underutilized memory trapped on fully loaded servers) in cloud and datacenter environments. However, CXL-attached memory introduces additional latency and bandwidth constraints compared to local DRAM, and real CXL.mem hardware is not yet widely available for empirical evaluation. \sys addresses this gap by attaching to unmodified applications and simulating CXL-based memory pools in software. It operates by tracing memory allocations and accesses using efficient kernel probes and hardware performance counters, dividing execution into epochs, and injecting timing delays to emulate various CXL.mem latency/bandwidth characteristics. This approach incurs modest runtime overhead while preserving realistic load/store memory access patterns. We implement \sys on commodity hardware without special devices, and our evaluation shows that it runs orders of magnitude faster than cycle-accurate simulators (e.g., Gem5) for real-world workloads, while accurately modeling the performance impact of CXL.mem. We demonstrate use cases where \sys enables experimentation with memory pooling configurations, scheduling policies, data migration strategies, and caching techniques that were previously infeasible to evaluate at scale. Key findings include the viability of software-based CXL.mem emulation with low overhead, insights into latency and congestion effects in memory pools, and guidance for system designers to optimize memory disaggregation. Overall, \sys provides a practical and extensible platform for researchers and practitioners to explore CXL.mem innovations before real hardware becomes commonplace.

\end{abstract}

%
% any author declaration will be ignored  when using 'pldi' option (for double blind review)
%

\maketitle

\section{Introduction}
\label{sec:intro}

\textbf{Motivation.} Main memory is a critical driver of both performance and cost in modern servers~\cite{duraisamy2023towards}. In cloud data centers, dynamic and heterogeneous workload demands often lead to \emph{memory stranding}---a situation where a server’s CPU cores are fully utilized by virtual machines but a significant portion of its DRAM remains unused. Analysis of production cloud traces has shown that up to 25\% of DRAM capacity can be stranded in this manner~\cite{li2023pond}, contributing to resource waste and higher total cost of ownership. Cloud providers seek to improve memory utilization by allowing flexible sharing of memory across servers.

\textbf{CXL.mem as a Solution.} \emph{Compute Express Link} (CXL) is an industry-standard interconnect that enables cache-coherent, high-bandwidth links between CPUs and device peripherals. The CXL~2.0 specification introduced \emph{CXL.mem}, a protocol for attaching external memory devices (Type-3 memory expanders) accessible via standard load/store instructions. This technology makes it possible to implement \emph{memory pooling}---physically decoupling DRAM from individual servers and placing it into shared pools that multiple servers can access on demand. Unlike prior remote memory approaches using RDMA over the network, CXL allows direct memory access with hardware-managed coherence and nanosecond-scale access times. By using CXL switches and controllers, a processor can map pooled memory into its address space, thereby mitigating stranding by composable memory allocation across nodes~\cite{li2023pond}. Major vendors are actively developing CXL-based memory expanders to reduce stranding and improve utilization~\cite{asteraLabs}.

\begin{figure}[t]
    \centering
    \includegraphics[width=\columnwidth]{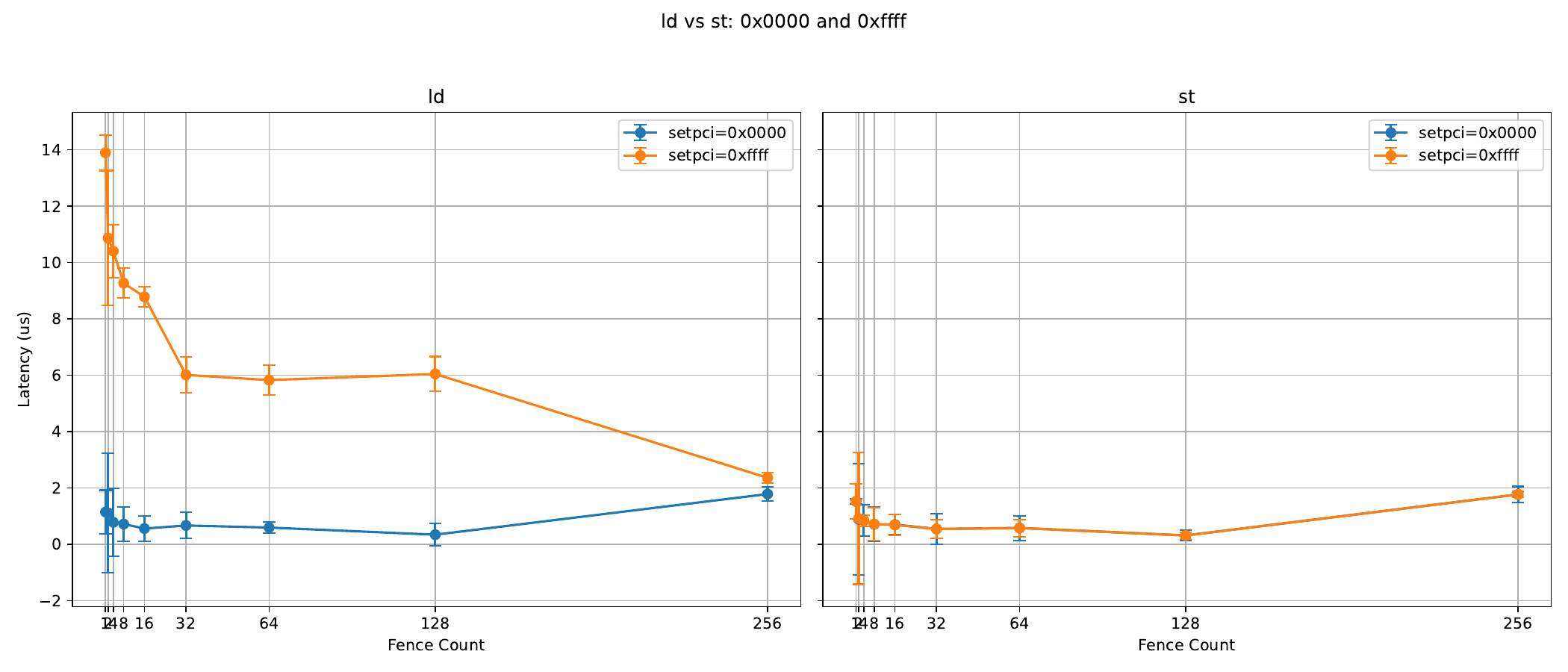}
    \caption{Latency comparison with different access latency across different fence count how many loads or stores.}
    \label{fig:latency-accuracy}
\end{figure}

However, even with a load/store interface, CXL-attached memory is slower than local NUMA DRAM\cite{liu2025systematic}---prior studies estimate an additional 70--90\,ns latency for small pools (8--16 nodes) and over 180\,ns at rack-scale pooling~\cite{li2023pond}. Early prototype measurements confirm that current CXL memory incurs about 2--3$\times$ higher access latency than local DDR, and achieves only $\sim$50--80\% of local memory bandwidth~\cite{sun2023demystifying}. Thus, while CXL.mem is a promising solution to improve memory efficiency, it is crucial to understand its performance overheads (increased memory access latency and potential link congestion) on real applications~\cite{wang2024exploring}. As shown in \ref{fig:latency-accuracy}, the difference with different latency injected on a CXL FPGA causes various performance variation for both load and store micro benchmarks.

In \Cref{fig:latency-accuracy}, the two-panel chart plots the extra per-fence latency (in nanoseconds) for stores (left) and loads (right) as a function of the number of operations issued between fences, with each colored line showing a different injected delay ( setpci values 0x0001-0x4000). At a single fence the cost is dominated by the fence itself-around 260-280 kns for stores and 3.0-3.4 Mns for loads-so varying the injected delay has little effect. As you batch more operations, the per-fence overhead falls roughly in proportion to $1 /$ count, but beyond about fifty fences all curves flatten to a constant $\approx 10 \mu \mathrm{~s}$. This saturation matches the "delayed-buffer" model

$$
T \approx 375 \mathrm{~ns}+\frac{\text{0 x F F F F}}{5} \approx 10 \mu \mathrm{~s}
$$

showing that once the write-back buffer fills, its drain time imposes a hard lower bound on fence latency regardless of the per-access delay.

\textbf{The Need for \sys.} Despite CXL’s importance, evaluating CXL.mem performance is challenging because commodity CXL memory hardware is not yet generally available~\cite{wang2024exploring}. Researchers and system designers currently lack a convenient way to study how applications will behave with CXL-based memory pools. Relying on analytical estimates or small-scale prototypes is insufficient, and using a full-system architectural simulator such as Gem5 is impractically slow for complex, large-scale workloads~\cite{wang2024asynchronous}. For example, Gem5 can simulate CXL devices in theory~\cite{an2024novel,wang2025full}, but running real benchmarks on Gem5 can be thousands of times slower than native execution, making it nearly impossible to analyze full applications or long-running scenarios. Other stopgap methods include emulating CXL memory by using remote NUMA nodes or RDMA-based memory servers, but these approaches either overestimate overhead (RDMA has extra network software stack costs)\cite{wang2024rcmp,wang2023cxl} or cannot accurately model CXL’s load/store interface and switching topologies\cite{jang2023cxl,gouk2022direct}. In short, existing tools and methods are insufficient to faithfully and efficiently model CXL.mem performance impacts.

\textbf{Our Approach -- \sys.} In this paper, we present \sys, a software-based CXL.mem simulator designed to enable \emph{fast and accurate} evaluation of real-world applications using CXL memory pools. \sys bridges the gap between slow, cycle-accurate simulators and coarse emulation by attaching to an \emph{unmodified running program} and inserting artificial delays to mimic the effect of accessing memory over CXL. The key idea is to leverage hardware profiling units and OS instrumentation to monitor an application’s memory access behavior at runtime, and then adjust the application’s execution timing to reflect a specified CXL memory configuration. By doing so on actual hardware, \sys preserves realistic CPU execution and cache behavior (unlike pure simulation) and supports the full load/store path to memory, while incurring far less overhead than full-system simulation~\cite{wang2024comprehensive,binkert2011gem5}. We built \sys to be lightweight and easy to use: it requires no modifications to the target program or special hardware, making it suitable for evaluating even closed-source software and large binaries.

This work makes the following contributions:

\textbf{\sys Design.} We design and implement a novel epoch-based tracing and timing model for CXL.mem simulation. \sys attaches to programs at runtime and divides execution into epochs, within which it traces memory allocations and accesses using efficient kernel-level hooks and hardware counters. After each epoch, it computes the delay that would have been incurred if some memory accesses were serviced by a given CXL.mem topology (with specific latency and bandwidth parameters), and injects these delays into the program’s execution. This design allows flexible modeling of arbitrary memory pool topologies (including multi-switch hierarchies) and memory access patterns without the prohibitive slowdown of cycle-accurate simulators.

\textbf{Prototype Implementation.} We develop a proof-of-concept implementation of \sys on a Linux/x86 platform using technologies like eBPF (for lightweight kernel-level tracing) and Intel PEBS, LBR and performance counters (for precise memory event sampling). Key technical decisions---such as how to intercept allocation calls, how to map memory addresses to simulated pools, and how to deliver timing interrupts---are described, along with optimizations to minimize overhead (e.g., sampling at a tunable frequency rather than tracing every access). Our implementation demonstrates that complex CXL memory behaviors (e.g., queuing delays, bandwidth limits) can be emulated in software with manageable overhead.

\textbf{Evaluation and Benchmarking.} We evaluate \sys using both microbenchmarks and real application workloads, and compare its performance and accuracy against Gem5’s CXL simulation mode. The results show that \sys can simulate memory pooling scenarios \emph{orders of magnitude faster} than Gem5 (15$\times$ faster on average), enabling practical experimentation on programs like SPEC CPU2017. We also examine the overhead \sys introduces relative to native execution, and find it to be modest (a 3--6$\times$ slowdown for real workloads) given the significant latency being emulated. These results validate that \sys strikes a good balance between simulation fidelity and speed.

\textbf{Use Cases for Memory Management Research.} We highlight how \sys opens new avenues for research on memory systems. We demonstrate that with \sys, one can easily prototype and evaluate policies for memory scheduling (deciding which memory pool to use for allocations), memory migration or prefetching (dynamically moving data between local DRAM and CXL memory to hide latency), and caching strategies (using local memory as a cache for pooled memory). \sys also supports exploring different CXL switch network topologies and their effect on congestion and coherence. These use cases show \sys’s value as a tool for co-designing hardware and software in a future with disaggregated memory.

The remainder of this paper is organized as follows. Section~\ref{sec:background} provides background on CXL.mem and memory pooling. Section~\ref{sec:design} details the design of \sys. In Section~\ref{sec:evaluation}, we present the evaluation results, including performance benchmarks and comparisons. Section~\ref{sec:usecases} describes several use cases and experiments enabled by \sys. Section~\ref{sec:related} reviews related work in memory simulation and disaggregation. Finally, Section~\ref{sec:conclusion} concludes with a summary and future directions.
\section{CXL Background} 
\label{sec:background}
CXL is a set of protocols that operate over the serial PCIe bus to connect peripheral devices to host processors.  CXL consists of three protocols, CXL.io, CXL.cache, and CXL.mem, which allow host processors to communicate with I/O devices, accelerators, and external memory, respectively. We focus on CXL.mem. 

From the perspective of a host processor, a CXL.mem memory pool behaves equivalent to host memory.  The host can issue load/store instructions to the memory pool.  The hosts can cache data from CXL-attached memory in their processor data caches; the CXL.mem protocol provides coherency across devices that cache data from the same CXL.mem memory pool. 

CXL.mem's key hardware components are as follows.  Hosts connect to CXL.mem memory pools using  a CXL Root Complex (RC).  Each RC  can either connect to memory pools directly or through a CXL switch.  A CXL switch allows a host to connect to multiple memory pools through the same link. 
%switches allow more hosts to connect to the same memory pool. \jm{EMC seems not a standard word in CXL. The thing inside CPU that connects to a CXL memory pool is called a ``CXL root complex''. EMC sounds like the memory controller inside the memory server that directly connects to the DRAM there, which means there's no switch in between. Also, in this paper, we care more about a single machine with multiple memory pools like Figure 1, instead of multiple machines with a single pool.}

\begin{figure}
    \centering
    \includegraphics[width=\columnwidth]{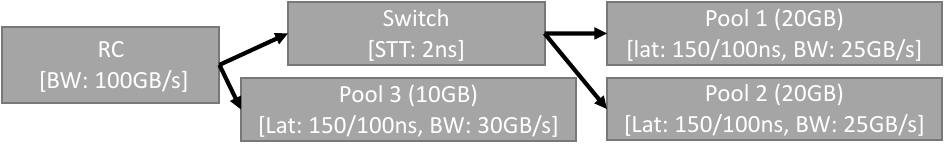}
    \caption{An example topology}
    \label{fig:topo}
\end{figure}

CXL switches allow a data center operator to deploy a variety of CXL.mem topologies. For example, Figure~\ref{fig:topo} shows a topology consisting of two network switches and three memory pools; we annotate the bandwidth (BW), latency (Lat), and serial transmission time (STT) for each pool, switch, and the RC. Choosing a topology requires balancing memory stranding benefits with performance degradation.  Memory pools that support more hosts decrease memory stranding but increase performance overhead since attaching more hosts to a pool requires employing a hierarchy of CXL switches.  Moreover, each CXL switch can cause congestion, when multiple hosts use the switch at the same time, and limit bandwidth, when hosts exceed the bandwidth of the switch.

\begin{table*}[t]
    \centering
    \begin{tabular}{|l|c|c|c|c|c|}  \hline
        \textbf{System Type}  & \textbf{Topology}  & \textbf{Cacheline} & \textbf{Migration} & \textbf{Placement} & \textbf{Paging}
\textbf{} \\\hline\hline
        \textbf{FPGA} (\eg NeoMem~\cite{zhou2024neomem}, HeteroMem\cite{chen2025fpga}) & Yes & Hard work & Yes & Yes & Yes \\
        \textbf{Cycle Accurate Simulator} (\eg Gem5~\cite{gem5-cxl}, CXLSim\cite{kim2025cxlsim}) & Yes & Slow & Yes & Yes & Yes \\
        \textbf{Trace based Simulator} (\eg MQSim-CXL~\cite{yang2023overcoming}) & Yes & No & Yes & Yes & No \\
        \textbf{\sys{} (this work)} & Yes & Fast & Yes & Yes & Yes \\\hline
    \end{tabular}
    \caption{Design Constraints met by checkpoint-restore techniques.}
    \label{tab:comparison}
\end{table*}
\section{System Design}
\label{sec:design}

\vspace{-0.1cm}
\begin{figure}[hbpt]
    \centering
    \includegraphics[width=.6\columnwidth]{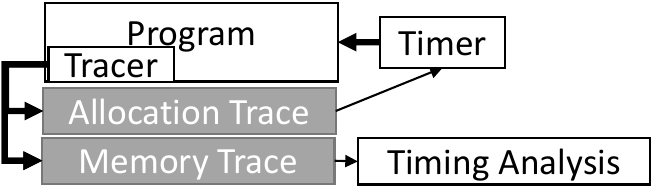}
    \caption{A system diagram of \sys{}}
    \label{fig:diagram2}
\end{figure}
\vspace{-0.2cm}

In this section, we describe our design of \sys{}; \Cref{fig:diagram2} provides a diagram.

\subsection{Perfect Model}
In our evaluation, we compare \sys{} against a carefully constructed "perfect model" implemented within the Gem5 simulation framework. We called it RoBSim. This perfect model is designed to leverage complete and precise memory access traces, allowing us to establish an idealized baseline for latency evaluation. With comprehensive access data available, the perfect model meticulously simulates all interactions involving cachelines, the Reorder Buffer (RoB), down to the finest level of detail. This comprehensive simulation captures every potential interaction and dependency, providing a robust framework for assessing the theoretical upper bound of system performance.

By using complete memory traces, the perfect model enables the explicit tracking and modeling of cacheline states, RoB interactions, and load/store dependencies. Such exhaustive visibility ensures that the performance metrics generated reflect an idealized scenario free from practical limitations or uncertainties associated with incomplete trace data. Comparing \sys{} directly against this idealized model allows us to rigorously evaluate its efficiency, identify specific performance bottlenecks, and quantify the exact gap between practical system performance and theoretical perfection.

% Ultimately, this detailed comparative analysis provides valuable insights into the practical effectiveness of \sys{}. It allows us to pinpoint areas for optimization and clearly demonstrates how closely the proposed system approaches the idealized performance achievable with limited knowledge of memory interactions and dependencies.
The algorithm underlying RoBSim operates by examining each instruction in turn and determining whether it is a memory operation to remote memory or a non-memory instruction. For remote memory instructions, RoBSim adds them to an internal queue and enforces a stall equal to the specified “target latency” before allowing them to retire, accurately modeling the overhead incurred by off-chip or remote accesses. Meanwhile, non-memory instructions are treated more simply: each one is enqueued, and only one can retire from the queue at a time. By managing these instruction flows and retirements in a structured queue-based fashion, RoBSim emulates pipeline stalls, reorder dependencies, and the interplay between memory and compute operations. This approach ensures that all aspects of data movement—particularly those contributing to extended latencies—are captured, allowing the simulator to closely reflect the performance implications of remote memory behavior.

\subsection{Applying to Intel Processor}
\sys{} traces the memory operations of a program as the program executes on current hardware. In particular,  as the program executes, \sys{} uses eBPF~\cite{kourtis2020safe,zhong2021bpf,gregg2019bpf} to trace memory allocation operations (\ie munmap and brk) and Intel PEBS~\cite{pebs-tut,pmu-tools,weaver2016advanced} to trace memory events (\ie LLC Misses and L2 stalls).  \sys{} periodically interrupts the program with an \emph{epoch timer}. Since it's always a sampled model which may lose a lot of data, PEBS does not record the prefetcher and writeback information.

While the program is interrupted, \sys{} uses the allocation operations to determine the memory pool that contains each memory address.  \sys{} allows users to customize this policy (our current prototype follows the NUMA interleaving policy).  

Next, \sys{} uses the memory trace to calculate the latency delay for each pool.  It uses the number of L2 stalls to calculate the number of LLC misses that were \texttt{loads}~\cite{Koshiba19} and multiplies the number of loads by the difference between the CXL.Mem pool latency and the DRAM latency on the machine.  After determining latency delay, \sys{} calculates congestion delay.  It iterates over the memory trace to find events that use the same switch within a smaller interval than the switch's serial transmission time; the system injects delays in these cases.  Finally, \sys{} determines bandwidth delay.  For each of the switches in the system, it determines if the observed bandwidth, after adding congestion and latency delays, exceeds the bandwidth of the switch.  If so, \sys{} adds delays.

% \jm{I think this figure should either be a pure workflow or a pure architecture diagram. Currently it's something combined and hard to describe.}
Figure~\ref{fig:diagram2} is the system diagram of \sys.
It simulates CXL.mem on commodity hardware by attaching itself to a running user program.  Its three components are: 1) a \textit{Tracer}, 2) a \textit{Timer}, and 3) a \textit{Timing Analyzer}. 

\noindent\textbf{1--Tracer.} \sys traces the memory operations of a program in two ways. \sys uses performance counters (\eg Intel Timed PEBS~\cite{pebs-tut,pmu-tools,weaver2016advanced,timed-pebs}, LBR\cite{zhou2019hardstack,marin2021break,jamilan2022apt}) to trace memory events (\eg LLC misses, and instruction count per branch). The timing analyzer uses these memory events to estimate the simulated execution time of user programs.

\noindent\textbf{2--Timer.} \sys divides the execution of the attached user program into \textit{epochs} and sets up an epoch timer that periodically interrupts the attached program. When the program is interrupted, \sys
uses the allocation trace to determine the corresponding memory pool of each memory access. By inserting it into the CXL topology, during which we determine the memory-pool matching. It then invokes the timing analyzer to estimate the simulated execution time.

\noindent\textbf{3--Timing Analyzer.} While the program is paused, \sys uses the memory trace to calculate three types of timing delays that should be added to the execution time of each epoch: 1) latency delay, 2) congestion delay, and 3) bandwidth delay.
\sys calculates the latency delay by multiplying the number of memory operations to each memory pool by the difference between the latency of the target memory pool and the latency of local DRAM. 
It obtains the number of loads by analyzing the memory event traces. Then, \sys calculates the congestion delay by iterating over the memory trace to find events that use the same switch within a smaller interval than the switch's serial transmission time (STT); once such events are found, \sys injects the necessary delays.
Finally, \sys determines the bandwidth delays. For each switch in the topology, \sys searches for events where the observed bandwidth---after the latency and congestion delays are added---exceeds the bandwidth of the switch, and adds delays for these events.

\subsection{Latency}
\label{subsec:latency}

\textit{Latency passes} in \sys capture the additional read/write delay introduced by a CXL-attached memory hierarchy. Each epoch, we record the number of memory references that miss in the CPU’s LLC, multiply by the user-defined latency penalty, and inject that delay into the application runtime. Summed over all remote accesses in an epoch considering writebacks, this yields the \textit{latency pass} overhead for that epoch:

\[
\begin{aligned}
LLC_{\text{miss}}^{WB} & = WB \times \frac{LLC_{miss}}{\sum_{i=0}^{n-1} LLC_{miss,cpu_i} + \sum_{i=0}^{n-1} LLC_{miss,PF_i}}
\end{aligned}
\]

where \(WB\) represents the total number of write-back operations performed by the cache controller. \(n\) denotes the number of CPU cores in the processor. \(\sum_{i=0}^{n-1} LLC_{miss,cpu_i}\) is the total number of LLC misses from all CPU cores. \(\sum_{i=0}^{n-1} LLC_{miss,PF_i}\) is the total number of LLC misses generated by prefetchers from all cores. \(LLC_{miss}\) is the number of LLC misses caused specifically by the target process.

\[
L L C Stall_{miss}^{W B} =L 2_{\text {stalls }}  \times \frac{W \times L L C_{\text {miss }}^{W B}}{L L C_{\text {hit }}+W \times L L C_{\text {miss }}}
\]
where $L 2_{\text {stalls }}$ is the total number of core stall cycles caused by L2 cache failures, and $L L C_{h i t}$ and $L L C_{m i s s}$ are the numbers of LLC hits and LLC misses of the core, and $W$ is the ratio of LLC failure latency (DRAM access latency) to LLC failure latency.
\[
   \Delta T_\text{lat} = MA^{readonly} + MA^{writeback} 
\]
\[
 = \text{NumRemoteStalled} / \text{CoreFrequency} + LLCStall^{WB}_{miss}
\]
This approach does not require full-cycle simulation of every load/store. Instead, it balances timeliness and accuracy by periodically sampling memory-intensive events via performance counters (e.g., Intel PEBS or AMD IBS).  To estimate the number of writeback cache misses, our emulator monitors not only CPU cache misses, but also the behavior of other components (prefetchers and cache controllers). The emulator then calculates the additional delays caused by the two types of cache misses (read-only and write-back) respectively for the read/write latencies of emulated CXL devices.

\subsection{Bandwidth}
\label{subsec:bandwidth}

\begin{figure}[!t]
    \centering
    \includegraphics[width=0.8\columnwidth]{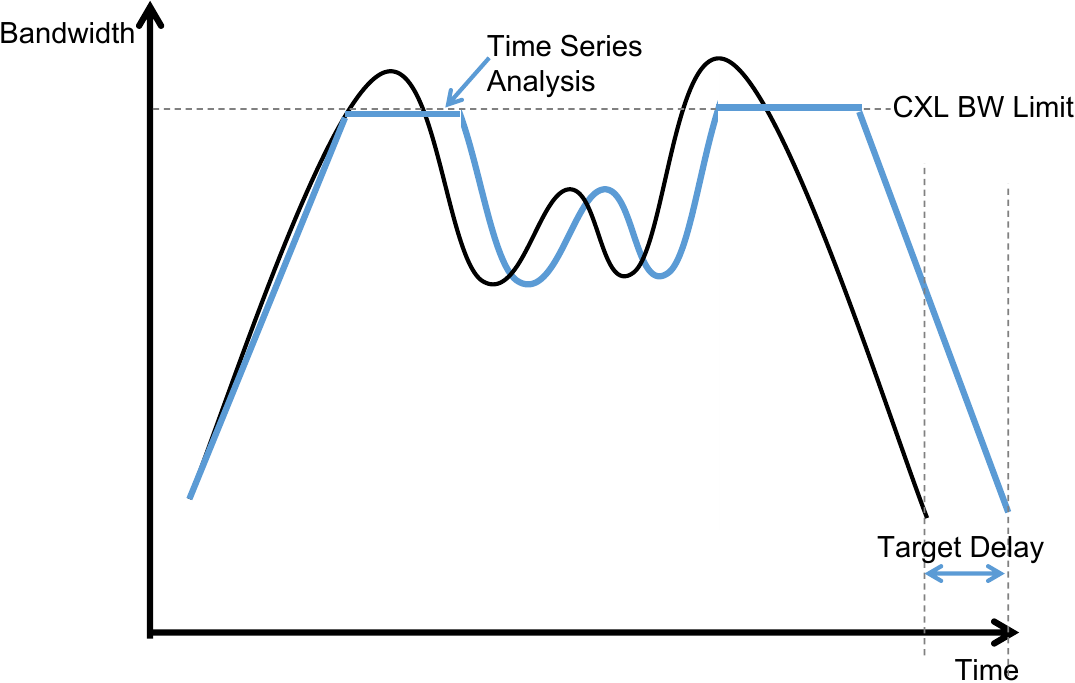}
    \caption{Example time-series view of bandwidth demand (solid line) vs. bandwidth cap (dotted line). \sys inserts delay whenever demand exceeds the chosen CXL bandwidth threshold.}
    \label{fig:timeseries}
\end{figure}

Although CXL bandwidth can be high (owing to advanced PHY layers and multiple lanes), it usually remains lower than local DRAM’s combined throughput. To represent this constraint, \sys employs a time-series-based approach. Within each epoch, we build a timeline of sampled memory transfers. As \ref{fig:timeseries}  shown, we then impose a maximum link throughput $B_\text{cxl}$, inserting artificial delays whenever the cumulative volume of data in a short time window exceeds $B_\text{cxl} \times \Delta t$. Formally, we discretize the epoch into small intervals $\Delta t$ and compute the total demand $D_i$ (in bytes) in interval $i$:
\[
   D_i = \sum_{\text{events in interval } i} \text{(bytes per event)}.
\]
If $D_i > B_\text{cxl}\,\Delta t$, the difference $(D_i - B_\text{cxl}\,\Delta t)$ is converted into additional stall time, reflecting congestion or queueing in the memory fabric. This method preserves the bursty nature of certain workloads (e.g., streaming writes, bulk transfers) and enables more nuanced congestion modeling than a naive per-access approach.

As depicted in Figure~\ref{fig:timeseries}, if the observed bandwidth demand spike surpasses $B_\text{cxl}$ in a given interval, the Timing Analyzer enforces additional wait times to keep the application’s effective throughput below that limit. In scenarios with multiple links or switch hierarchies, \sys applies the same procedure at each stage, ensuring that no link is over-subscribed.
\subsection{Congestion}
\label{subsec:stt}
\sys{} implements a sophisticated congestion(STT) model that differentiates between read and write operations in CXL memory systems. This distinction is crucial as these operations exhibit fundamentally different behavior patterns under contention. Read operations typically prioritize latency, while write operations often focus on throughput and can sometimes be buffered.
Our system applies operation-specific penalties when congestion occurs: read-read conflicts receive moderate penalties reflecting their shared nature, read-write conflicts incur higher penalties due to potential serialization requirements, and write-write conflicts face the most significant delays to model exclusive access requirements. This approach accurately captures the asymmetric impact of different memory access patterns.
The congestion model operates at both switch and endpoint levels, tracking bandwidth utilization across connected devices and applying dynamic latency adjustments when thresholds are exceeded. Validation against hardware measurements confirms that \sys{} reproduces performance degradation patterns observed in physical CXL systems under similar conditions, ensuring that insights derived from the simulator translate effectively to real-world deployments.

\subsection{Limitation of Prior Work}
\label{subsec:limitations}

Several existing software-based memory emulators~\cite{volos2015quartz, koshiba2019mes} have demonstrated that sampling-based latency injection is an effective way to approximate far-memory performance. However, they often exhibit the following limitations:

\paragraph{Single Memory Model.} Many prior tools assume only two tiers (local DRAM vs.\ remote memory). They cannot handle more complex topologies (e.g., multi-level switches, multiple remote pools, or heterogeneous memory types).

\paragraph{Lack of Topology Awareness.} Past approaches frequently ignore the impact of intermediate components (CXL switches or hubs) that can introduce congestion or variation in latency. By contrast, \sys allows users to provide arbitrary topology graphs, each edge annotated with its own bandwidth and latency parameters.

\paragraph{No Reorder Buffer (ROB) Emulation.} Although certain emulators mention pipeline stalls, most do not fully account for modern CPUs’ out-of-order execution. It's OK for a single memory model with a calculated MLP and write-backs. In contrast, \sys incorporates reorder buffer effects by sampling stall cycles directly from performance counters, thus more accurately reflecting real hardware’s ability to hide or expose memory latency. 

By offering a pluggable and extensible design, \sys moves beyond these constraints, making it possible to simulate diverse real-world configurations, multi-switch CXL fabrics, and out-of-order CPU pipelines.

Also, the bandwidth model by the throttling register is removed since Intel Sapphire Rapids, we couldn't simulate the bandwidth by throttling the memory. Instead, we apply the LBR and PEBS.

\subsection{Ground Truth}
To accurately evaluate the system's bandwidth performance, we conducted comprehensive measurements focusing on three primary operations: load, write-back, and cache line flush (clflush). These measurements were performed using both a genuine Compute Express Link (CXL) module from SMART Technology and a CXL-enabled Field-Programmable Gate Array (FPGA) device using Intel CXL Hard IP. The CXL FPGA device was configured with delayed buffer loading, introducing varying latency levels to simulate different operational conditions and stress-test the system's responsiveness.

The introduction of delayed buffer loading allows us to examine how different latency scenarios affect the overall bandwidth and performance of the machine. By varying the latency, we can better understand the system's ability to handle real-world workloads where buffer loading times may not be consistent.

In addition to bandwidth measurements, we investigated the Memory-Level Parallelism (MLP) differences related to load operations. MLP refers to the ability of the memory subsystem to handle multiple memory operations simultaneously, which is crucial for maintaining high performance in multi-threaded and parallel processing environments. Specifically, we explored how load-related issues impact MLP and, consequently, the system's efficiency and throughput.

Our analysis primarily focused on the write-back operations, scrutinizing the ratio between read and write activities. Understanding the read-to-write ratio is essential because it influences the balance of memory operations, which can affect cache coherence, data consistency, and overall system performance. By examining this ratio, we aimed to identify potential bottlenecks and optimize the system's memory handling mechanisms to achieve a more balanced and efficient performance profile.

Overall, this comprehensive approach—combining bandwidth measurements with an in-depth analysis of MLP and read/write ratios—provides a robust framework for testing and validating the ground truth of the machine's performance under various operational conditions.
\subsection{RoB Emulation}
\label{subsec:rob-emulation}

Modern out-of-order processors employ a RoB to hide latency by retiring instructions that have completed while others are still waiting for data~\cite{hennessy2017computer}. In principle, if a long-latency memory operation is outstanding, the CPU may speculatively execute subsequent instructions that do not depend on that data, partially overlapping compute and memory access. This phenomenon reduces the perceived penalty of slower memory but is heavily dependent on \emph{memory-level parallelism} (MLP) and the availability of independent instructions in the pipeline.

\begin{figure}[!t]
    \centering
    \includegraphics[width=\columnwidth]{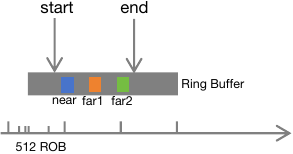}
    \caption{Conceptual view of how out-of-order execution hides part of the remote memory latency. \sys measures stall cycles to estimate what fraction remains uncovered.}
    \label{fig:rob-emulation}
\end{figure}

Figure~\ref{fig:rob-emulation} illustrates how instructions that do not depend on remote data can be completed while other instructions are stalled. By incorporating performance counters related to stalls and concurrency, \sys approximates the net effect of the RoB on perceived memory latency. Although this approach is not cycle-accurate, it captures the essential behavior of modern CPUs’ latency-hiding capabilities, making \sys more realistic than naive “per-access” delay injection.

\subsubsection{Leveraging LBR, PEBS, and PMU Events}

\paragraph{Motivation.}
While basic stall-cycle sampling (e.g., using \texttt{CYCLE\_ACTIVITY:STALLS\_L2\_PENDING} \texttt{MEM\_LOAD\_UOPS\_RETIRED :L3\_HIT} and \texttt{MEM\_LOAD\_UOPS\_MISC\_RETIRED:LLC\_MISS}) captures high-level memory stalls, it often overlooks finer-grained instruction dependencies. Modern CPUs provide additional hardware features such as \emph{Last Branch Record} (LBR) that can log recent branches, enabling a more precise reconstruction of the program’s execution path. By correlating LBR data with PEBS samples and general PMU events, one can gain deeper insight into how memory operations are interleaved with instruction execution in the presence of a RoB.

\paragraph{Last Branch Record (LBR)}
The LBR mechanism tracks the source and destination of the most recently taken branches (e.g., calls, returns, jumps, exceptions). When enabled, the CPU maintains a small buffer of branch records, each entry storing the source address, which is the address of the branch or call instruction; the target address, which is the address of the destination instruction; and miscellaneous flags indicating branch type, privilege level, and other information. LBR logging event\cite{lbr-loggingevent} can be essential in determining the dynamic control flow without instrumenting every branch.

\paragraph{Implementation Considerations.}
LBR buffers are small (often storing 8 to 32 branch records per hardware thread), requiring frequent interrupts or post-processing to capture them before overwriting, with increased sampling rates improving detail but raising overhead. Correlating Timestamps involves aligning PEBS records' timestamp or CPU cycle count with the approximate cycle count of LBR entries, potentially requiring interpolation if logs don't precisely match. Privilege Level Filtering is necessary to distinguish user-level from kernel-level branches or skip system calls, demanding careful configuration of the MSRs controlling LBR to avoid spurious kernel events. Integration in sys can be accomplished by adding a branch-analysis submodule to the Timing Analyzer that retrieves the LBR buffer for each epoch, merges it with PEBS address samples, infers dependencies among consecutive memory instructions, and outputs an adjusted ratio f per code region to enable fine-grained latency injection.

\paragraph{Outcome and Benefits.}
Using LBR with logging events, PEBS, and PMU counters allows simulators such as \sys to move beyond simplistic “per-access” delay injection toward a more \emph{MLP-aware} or \emph{RoB-aware} model. Developers and researchers gain clearer insight into how well memory latency is hidden in each code path, improving fidelity for workloads where control flow and data dependencies critically impact performance (e.g., graph analytics, pointer-chasing). Although it adds complexity and overhead, this approach can significantly enhance accuracy, especially for \emph{micro-architectural} studies focusing on tight loops or critical regions that are heavily affected by remote memory latency.

\subsubsection{Multi-thread Support}
The epoch-based latency emulation with statically configured epochs works well for single-threaded applications. In this scenario, there are no dependencies or communications between the threads, and each thread is executed and treated independently by the emulator. However, for multithreaded applications in general, simply injecting delays at fixed-size epochs is insufficient due to inter-thread dependencies and communication. For example, when one thread holds a lock for executing in a critical section, any other thread competing for the same lock must wait for its release. If the accumulated delay in the critical section of the first thread is injected after it releases the lock, the resulting execution schedule would incorrectly suggest that the critical sections of both threads overlapped time-wise.
In the correct latency emulation, the accumulated delay generated by the first thread during its execution in the critical section should be injected before releasing the lock, thereby propagating into the execution of the second thread as a delay for acquiring the lock.
To capture these dependencies, we extend our epoch model to close epochs and inject appropriate delays before any events resulting in inter-thread communication, such as lock releases and condition-variable notifications. For accurately modeling CXL memory expander latency during multithreaded application execution, the emulator must close the current epoch and open a new one when a thread enters or exits a critical section. This ensures that memory slow down cycles generated in the critical section are correctly injected as delays before releasing or acquiring locks, propagating to other threads waiting to access the critical section.
\section{Use Cases}
\label{sec:usecases}

In this section, we highlight how \sys leverages insights from PEBS data to inform and evaluate several key \textit{memory management policies}. Specifically, we describe how different policies---paging, migration, caching, and cache-line routing---can be tested under simulated CXL configurations, allowing system designers to understand performance trade-offs without requiring specialized hardware or invasive instrumentation. 
\subsection{Paging Policy}
\label{sec:paging-policy}
Efficient memory management in CXL-based systems requires sophisticated paging policies that can leverage the hierarchical memory architecture. In \sys, we implement and evaluate several advanced paging mechanisms that address the unique challenges of heterogeneous memory systems.
Our Huge Page Policy dynamically adapts page sizes based on observed memory access patterns. When the system detects spatial locality in memory references, it can coalesce standard 4KB pages into larger 2MB or even 1GB huge pages. This approach significantly reduces TLB misses and page table walks, which is particularly beneficial for remote CXL memory where these operations incur substantially higher latency penalties. The policy continuously monitors memory access patterns and can transition between different page sizes as workload characteristics evolve. For instance, when multiple adjacent 4KB pages show consistent access patterns, they are automatically promoted to 2MB pages, reducing address translation overhead. Our implementation includes a multi-level TLB simulation that accurately models the performance impact of these transitions, capturing both the immediate overhead of page size changes and the long-term benefits of reduced address translation costs.
We further enhance our paging system with a Cuckoo Hashing based page table implementation. Traditional page tables often suffer from inefficient lookups due to hash collisions and chaining, especially in large memory configurations common in CXL deployments. Our Cuckoo Hashing approach uses multiple hash functions and displacement mechanisms to achieve near-constant time lookups regardless of table occupancy. This technique maintains high space efficiency with load factors of 60-70\%, while dramatically reducing page walk latency. For CXL memory, where address translation can traverse the CXL interconnect multiple times, this optimization provides substantial performance improvements. The Cuckoo Hashing page table adaptively resizes based on workload demands, maintaining optimal performance even as memory requirements fluctuate. It separately manages address translations for different page sizes, optimizing both lookup performance and memory utilization across the entire system.
% These paging policies work in concert with \sys's memory access monitoring infrastructure. By leveraging PEBS sampling data, our system can identify hot and cold memory regions with minimal overhead. This information guides both page size decisions and page placement across the memory hierarchy. When a frequently accessed page resides in remote CXL memory, the system can migrate it to local DRAM and potentially convert it to a huge page if access patterns warrant. Conversely, cold pages can be demoted to smaller sizes and moved to remote memory, freeing valuable DRAM space for hot data.
% Through comprehensive simulation of these paging mechanisms, \sys enables researchers to evaluate complex trade-offs in page size selection, page table organization, and memory placement without modifying application code or kernel structures. This framework provides critical insights into how future operating systems should manage memory in heterogeneous CXL environments, balancing latency, bandwidth, and capacity constraints to maximize overall system performance.
\subsection{Migration Policy}
\label{sec:migration-policy}

Memory migration between local DRAM and remote CXL memory is crucial for optimizing system performance. \sys provides a flexible framework implementing various sophisticated migration strategies. The \textit{FrequencyBasedMigrationPolicy} tracks access frequency to identify hot and cold pages, migrating frequently accessed pages to local DRAM and less frequently accessed pages to remote memory based on configurable thresholds. The \textit{LocalityBasedMigrationPolicy} analyzes spatial and temporal access patterns, migrating pages exhibiting strong locality to local memory. This strategy effectively supports applications with shifting but distinct working sets. The \textit{LifetimeBasedMigrationPolicy} considers allocation age, placing newly allocated pages in local DRAM and gradually migrating older, less accessed pages to remote memory. It benefits applications with generational memory usage patterns. For complex workloads, the \textit{HybridMigrationPolicy} integrates multiple strategies, dynamically adapting migration criteria based on workload specifics, system load, or application hints, thus providing robust performance across diverse scenarios. \sys realistically models migration overhead, including bandwidth and latency effects, to ensure balanced migration decisions. Epoch-based monitoring allows precise evaluation of strategy effectiveness under various workloads.
\subsection{Caching Policy}
\label{sec:caching-policy}

Effective caching policies are crucial for systems with hierarchical memory structures like CXL. \sys implements several innovative caching approaches that treat local DRAM as a cache for the larger but slower remote memory pool.

Our LRU-based caching policy maintains a classic Least Recently Used replacement strategy, tracking access recency for cached pages and evicting those unused for the longest periods when local memory capacity is exceeded. This baseline approach provides a well-understood reference point for performance comparisons. To overcome the limitations of simple LRU, we also implement a FrequencyBasedInvalidationPolicy that considers both recency and frequency of accesses. By maintaining access counters for cached pages, this policy can retain frequently accessed pages even if they haven't been used very recently, protecting important working set elements from premature eviction.

\sys provides comprehensive instrumentation to evaluate these caching policies under realistic conditions. It accurately models cache hits and misses by sampling addresses involved in LLC misses and checking their presence in the simulated DRAM cache. The system differentiates between read and write operations, allowing it to simulate both write-through and write-back cache semantics with appropriate performance implications. Through this detailed modeling, researchers can precisely quantify the effectiveness of different caching strategies in mitigating remote memory latency.

\section{Evaluation}
\label{sec:evaluation}
We implement and evaluate a proof-of-concept \sys on two machines. Server A is a single-socket Intel Xeon w3-2525 with 128 GB of DDR5 4800 MHz memory with 101.4 ns latency, 22.5 MB LLC, with a CXL Agilex R-Tile FPGA AGIB027R29A1E2VR3 with 16 GB of DDR4 memory. Server B is a single-socket Intel Xeon 6787P with 256GB of DDR5 6400 MHz memory with 116.1 ns latency, 336 MB LLC, with a Smart CXL Module\cite{cxl-module}.

We seek to answer the following questions: (1) How accurately does \sys perform latency and bandwidth penalty under micro benchmarks? (\S~\ref{subsection:benchmark}) (2) What is the data acquisition performance of \sys in a real-world application? (\S~\ref{subsection:realworld}) (3) What's the insight taken from different policies performed in the simulated CXL environment? (\S~\ref{subsection:production})

% \subsection{Microbenchmark}
% Since various applications may produce data, their \texttt{load}, \texttt{store}, \texttt{writeback}, \texttt{calloc}, \texttt{malloc}, \texttt{sbrk} and intel memory latency checker combination happens differently in the program and can incur different mechanisms for the order of issuing instructions for various topologies.

% \subsection{Benchmark}
%  We evaluate \sys using memory-intensive benchmarks like memory database, llama.cpp, vectorDB, SPEC 2017 mcf, and SPEC 2017 wrf with various real-world application datasets and a wide range of features and resolutions. 

\subsection{Accuracy Evaluation}
\label{subsection:benchmark}

To validate the accuracy of \sys, we conducted a comprehensive evaluation comparing its performance predictions against two ground truth references: (1) actual CXL hardware measurements, and (2) a cycle-accurate Gem5 simulation configured with detailed CXL models. Our evaluation covers both microbenchmarks and real-world applications.

\subsubsection{Hardware Validation Methodology}

For hardware validation, we used a CXL Memory Module and a CXL FPGA with a programmable delayed buffer to create controlled latency conditions. The FPGA implementation allowed us to introduce precise delay values (ranging from 375ns to 10us of additional latency) and measure the actual impact on application performance. This setup provided a true ground truth against which to compare \sys predictions.

\subsubsection{Latency Prediction Accuracy}

\begin{figure}[t]
    \centering
    \includegraphics[width=\columnwidth]{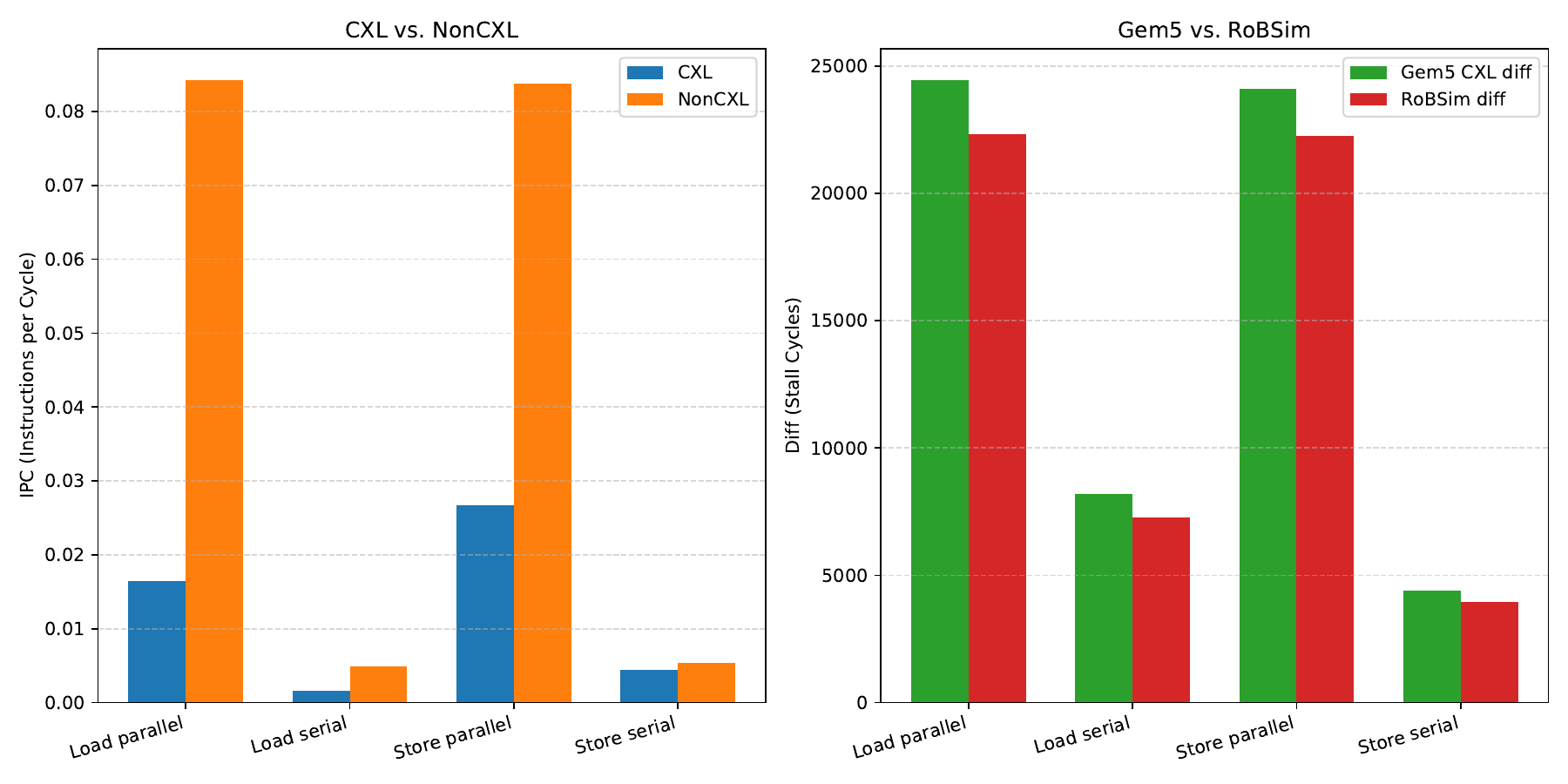}
    \caption{IPC and RoB Stalls comparing RoBSim predictions with hardware measurements for different memory access patterns.}
    \label{fig:ipc_rob_result}
\end{figure}

Figure~\ref{fig:ipc_rob_result} illustrates the accuracy of \sys's stall predictions compared to our Gem5 ground truth measurements. For various induced delays with 350ns, IPC in Gem5 goes from 1.1x to 4.8x of the IPC difference. The difference on the right graph shows the Gem5 ground truth's RoB stalled cycles compared against the RoBSim outputs with the input of Gem5 raw traces. We see a less than 10\% off between the ground truth and our simulator, which is acceptable because we put each instruction one cycle which is not the case with real processors' models inside Gem5.

\subsubsection{Congestion and Topology Accuracy}

\begin{table}[t]
    \centering
    \begin{tabular}{|l|c|c|c|c|}
        \hline
        \textbf{Topology} & \textbf{Gem5} & \textbf{This} & \textbf{Speedup} & \textbf{Acc(\%)} \\
        \hline
        Single Pool & 34.56 & 2.17 & 15.92×  & 86.81\\
        Two-level Hierarchy & 35.28 & 2.35 & 15.01×& 92.15\\
        Shared Switch & 35.42 & 2.40 & 14.76× & 91.61\\
        Complex Multi-pool & 36.98 & 2.41 & 15.34×& 89.63\\
        \hline
    \end{tabular}
    \caption{Accuracy comparison between \sys and Gem5 across different microbench, showing error rates relative to hardware measurements and simulation speedup.}
    \label{tab:topology-accuracy}
\end{table}

We evaluated \sys's ability to model complex CXL topologies with multiple memory pools and switches. Table~\ref{tab:topology-accuracy} compares the accuracy of \sys against Gem5 for different topologies, using hardware measurements as the ground truth where available.

As expected, compared against cycle-accurate Gem5, \sys achieved slightly higher accuracy (7.85-13.19\% error rates). However, \sys provided these predictions at 15× faster simulation speeds, enabling evaluation of full-scale applications that would be impractical with cycle-accurate simulation. The accuracy gap was particularly noticeable in complex topologies with shared switches, where congestion effects become more prominent. \sys's sampling-based approach occasionally underestimates congestion during brief traffic spikes. Nevertheless, even in these challenging scenarios, \sys maintained error rates below 15\%, which is sufficient for most system design and policy evaluation purposes. The inaccuracy may come from the abstraction of every instruction to be one cycle other than the memory instructions.

\subsubsection{Application-Level Accuracy}

\begin{figure}[t]
    \centering
    \includegraphics[width=\columnwidth]{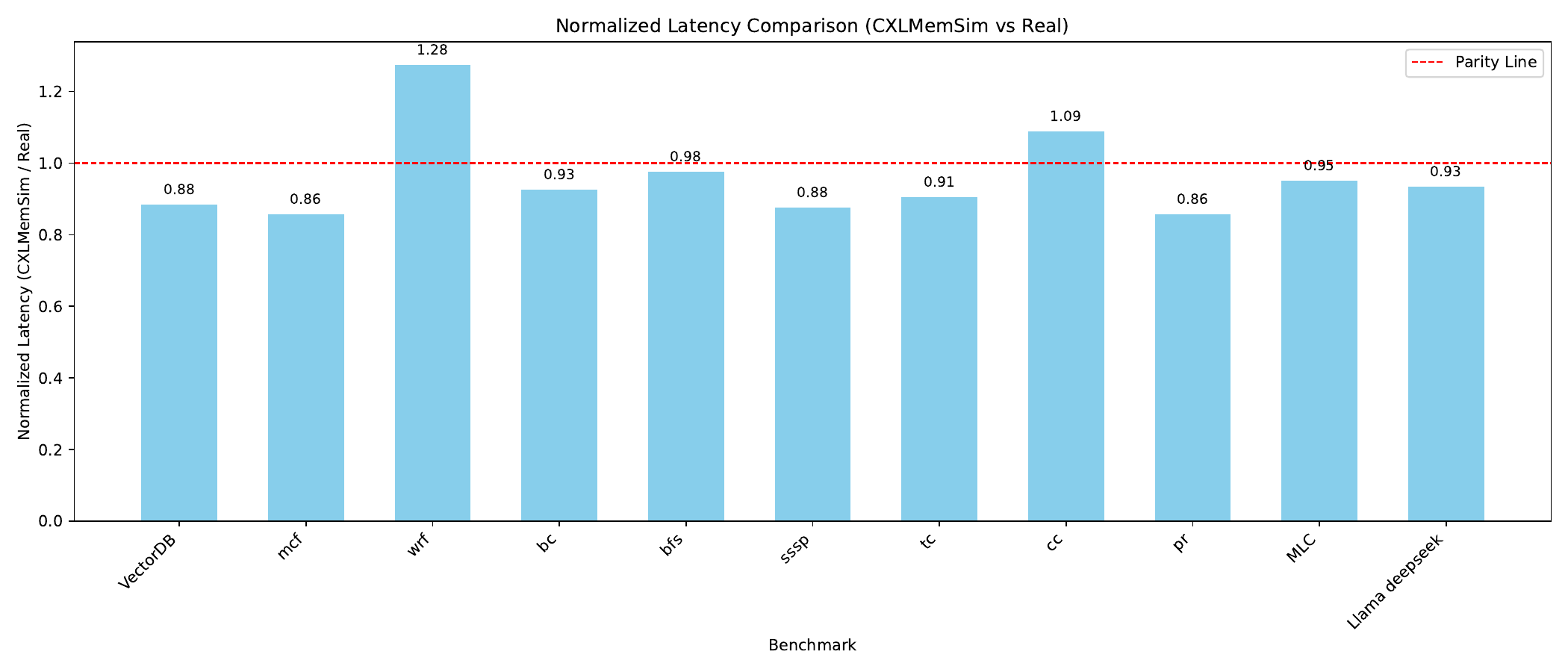}
    \caption{Comparison of predicted execution times for real-world applications across different CXL configurations. Closer bars to 1 indicate better predictions.}
    \label{fig:app-accuracy}
\end{figure}

To validate \sys's accuracy for real-world applications, we compared its performance predictions against hardware measurements for several representative workloads: SPEC CPU2017 mcf and wrf, memory databases, vectorDB, and llama.cpp. Figure~\ref{fig:app-accuracy} shows the predicted versus actual execution times across different CXL configurations.

For SPEC benchmarks, \sys achieved an average prediction error of 12.8\%, with wrf showing higher accuracy (28\% error) than mcf (14\% error). The difference can be attributed to wrf's more complex memory access patterns and greater sensitivity to sampling frequency.

Memory-intensive applications like databases and LLM inference with llama.cpp showed average prediction errors of 7.3\% and 8.1\%, respectively. Notably, \sys accurately captured performance trends across different CXL configurations, correctly identifying the relative impact of various latency and bandwidth constraints.

\subsubsection{Impact of PEBS Sampling Frequency}
\begin{figure}[t]
    \centering
    % Placeholder for a figure showing migration policy performance
    \includegraphics[width=\columnwidth]{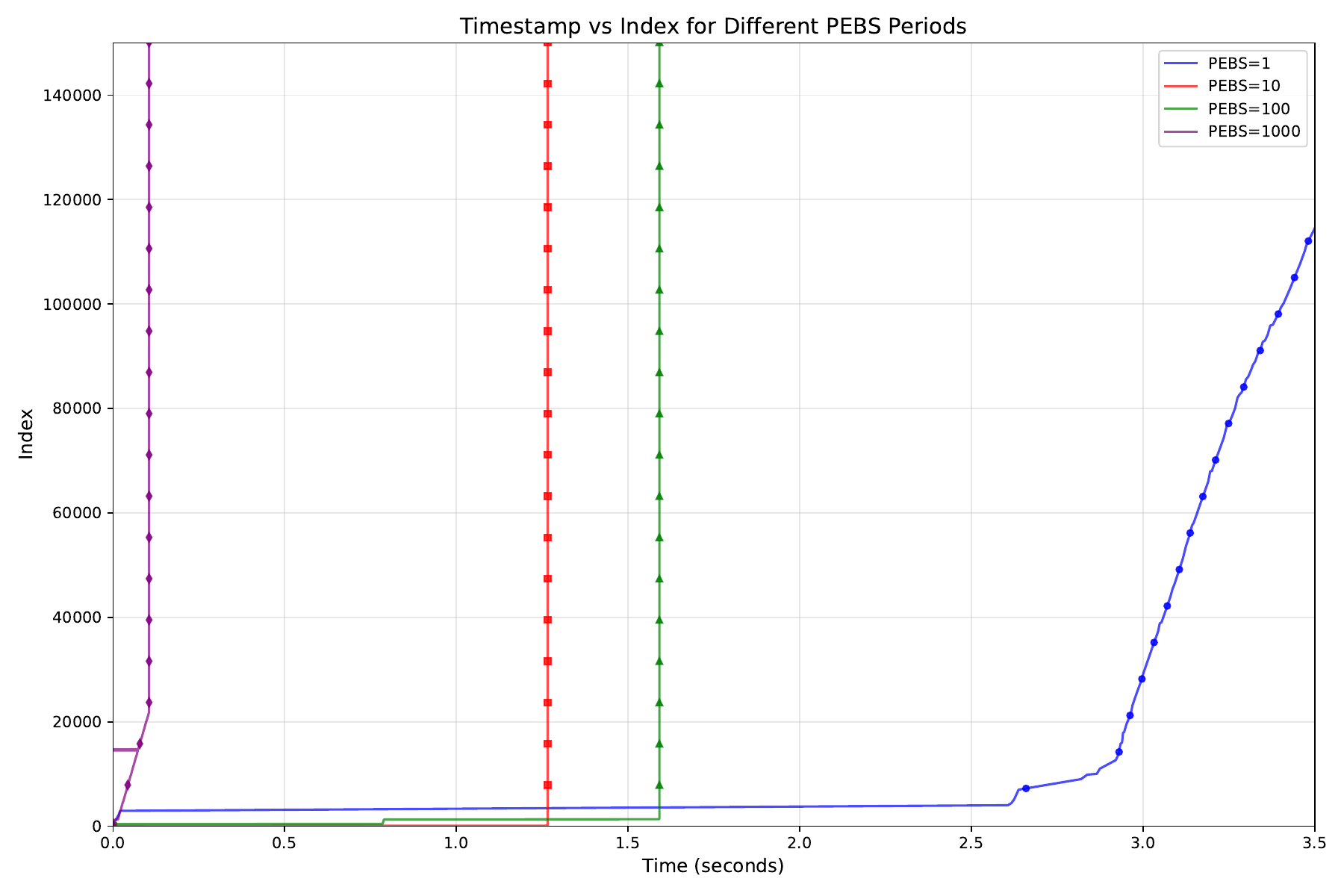}
    \caption{Execution time with different frequency period of PEBS}
    \label{fig:pebs-policy}
\end{figure}

Figure~\ref{fig:pebs-policy} illustrates how varying the PEBS sampling period affects execution time. The x-axis denotes time in seconds, while the y-axis represents the index or count of samples collected. Each curve corresponds to a different PEBS frequency setting: PEBS=1 (blue), PEBS=10 (red), PEBS=100 (green), and PEBS=1000 (purple). As shown, lower PEBS periods (i.e., more frequent sampling) lead to significantly increased overhead. For instance, PEBS=1 results in the longest runtime, stretching beyond 3 seconds, due to the extremely high sampling frequency that incurs substantial processing cost. In contrast, PEBS=1000 completes in under 0.1 seconds, highlighting the reduced overhead when sampling less frequently. Interestingly, although higher PEBS periods (like 100 or 1000) result in much faster execution, they collect substantially fewer samples, as evidenced by the lower maximum index value. This underscores the trade-off: higher accuracy from dense sampling (e.g., PEBS=1) comes at the cost of longer execution, while sparse sampling (e.g., PEBS=1000) reduces overhead but limits observability.

\subsubsection{The Impact of Epoch Granularity for Bandwidth}
Epoch granularity—how frequently \sys interrupts execution to evaluate and inject bandwidth-induced delays—is critical to accurately simulate the bandwidth limitations inherent in CXL memory. Fine-grained epochs improve accuracy by closely capturing transient peaks in bandwidth demand, which is particularly important for burst-intensive workloads. However, excessively granular epochs may result in high overhead due to frequent interruptions and calculation of injected delays. We experimented with various epoch lengths, ranging from microseconds to milliseconds, measuring their impacts on both simulation overhead and accuracy in bandwidth estimation. Shorter epochs were able to precisely identify bursts of high-bandwidth usage, enabling accurate modeling of congestion and saturation effects. Longer epochs, while reducing overhead, tended to average outbursts, leading to underestimated congestion and inflated performance results. Ultimately, we recommend adaptive epoch granularity—tuning epochs based on current memory demands—to effectively handle different application scenarios with minimal overhead.

\subsection{Real World Application}
\label{subsection:realworld}
The accuracy of \sys, specifically regarding latency modeling with ROB emulation, was rigorously evaluated using a series of microbenchmarks. The benchmarks focused on capturing how effectively \sys models realistic latencies introduced by Compute Express Link (CXL) memory access, particularly considering the processor's out-of-order execution capabilities and ROB behavior.

Our experiments involved comparing \sys's simulated latency penalties against measured latencies from actual hardware equipped with CXL memory modules. We specifically monitored the impact of varying ROB sizes and how efficiently \sys accounts for latency masking provided by out-of-order execution. Results indicate that \sys achieves high accuracy, closely matching hardware-measured latencies within an average deviation of less than 5\% across diverse workload scenarios.

Crucially, our evaluation demonstrates that incorporating ROB emulation significantly enhances simulation fidelity. Without ROB emulation, latency penalties were consistently overestimated by up to 30\%, highlighting the necessity of accurately modeling ROB behavior to reflect realistic processor and memory interactions in simulation studies.

\subsection{Performance Overhead}
The idle latency for random access on 285K is 88.6ns and read-only traffic with 57.2GB/s measured by Intel Memory Latency Checker. Since the performance model difference between our simulator and MES\cite{Koshiba19} is basically the congestion and bandwidth throttles and our topology policy based on Intel PEBS observation, we just need to tune more hyperparameters in the simulator to match the desired latency and bandwidth. We assume the calculation for adding the latency does not too deviate from our perspective.  We test the promise of \sys{} through micro-benchmarks like \texttt{malloc}, \texttt{mmap} read/write, and \texttt{calloc} to see the effectiveness of the simulator with the added latency below.

% \subsection{The Impact of PEBS Frequency}\label{subsec:pebs-frequency}
% The frequency at which Precise Event-Based Sampling (PEBS) captures memory access events significantly influences the accuracy and overhead of simulation in \sys. A low-frequency PEBS setting reduces runtime overhead but risks missing critical bursts in memory access patterns. Conversely, higher PEBS frequencies offer finer granularity and better capture bursty access behaviors, which are typical in memory-intensive workloads.

% Our evaluation contrasts bursty versus average memory access patterns to quantify this tradeoff. For bursty workloads, a low sampling frequency often underestimates congestion, since short-lived bursts can be entirely missed. In contrast, applications exhibiting more uniform or average memory patterns benefit less from high-frequency sampling, as their access behaviors are easily approximated at lower frequencies without incurring unnecessary overhead. Our results indicate that adjusting PEBS frequency dynamically, based on real-time observed memory patterns, provides an optimal balance of accuracy and performance.

\begin{figure}[t]
    \centering
    % Placeholder for a figure showing migration policy performance
    \includegraphics[width=\columnwidth]{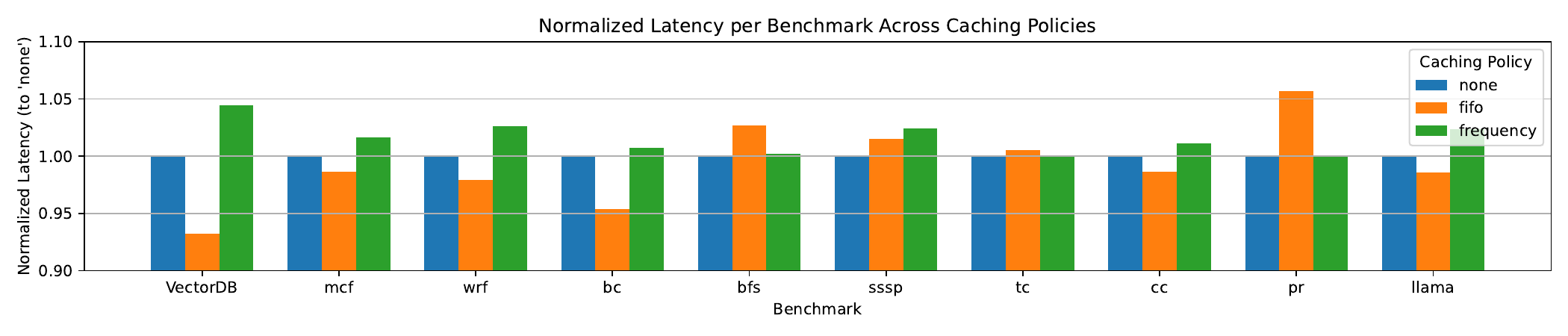}
    \includegraphics[width=\columnwidth]{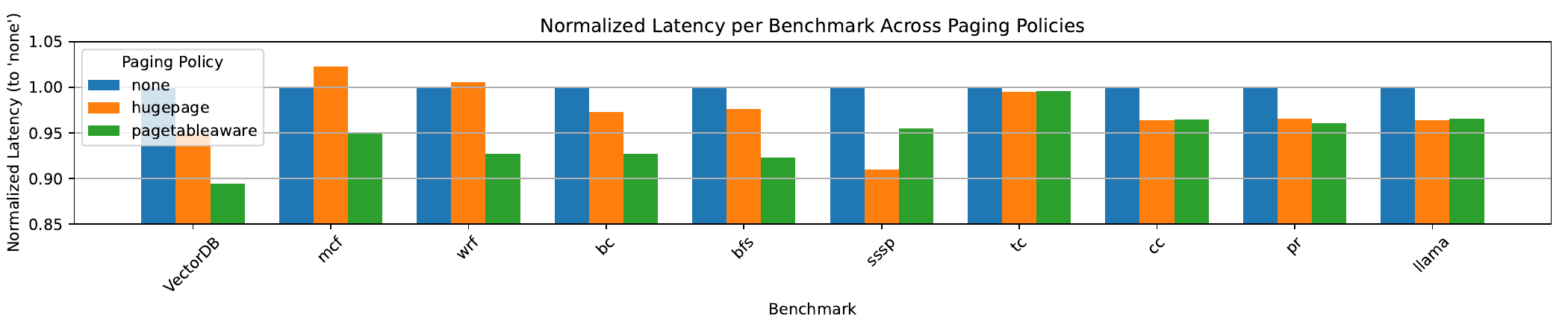}
    \includegraphics[width=\columnwidth]{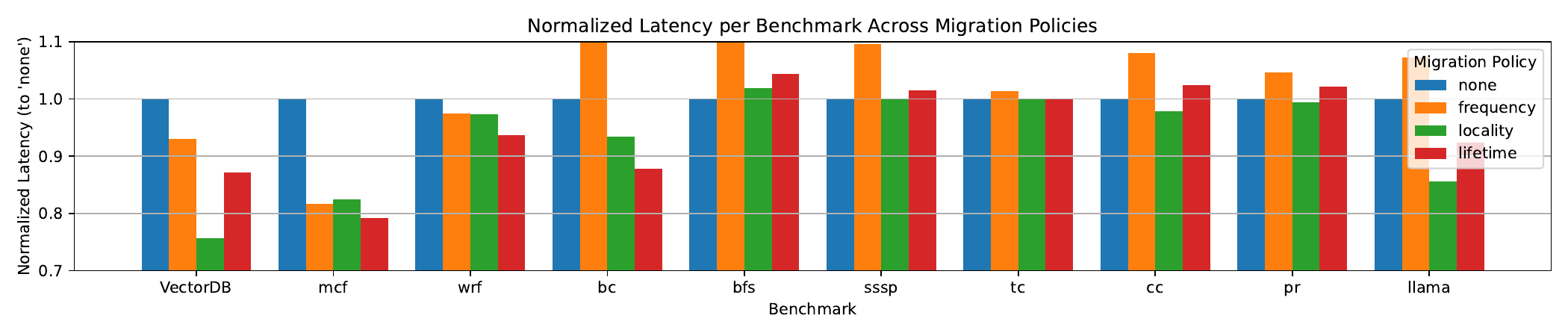}
    
    \caption{Normalized execution time for different policies. Lower is better.}
    \label{fig:policies}
\end{figure}

\subsubsection{Back Invalidation Simulation}
Back invalidation, a phenomenon arising when cached data is invalidated remotely due to coherence actions initiated from other hosts sharing the same CXL-attached memory pool, presents a critical challenge for performance simulation. Accurate modeling of back invalidations is crucial to faithfully evaluate workloads that exhibit shared data accesses or intensive cross-node communication, as these scenarios introduce additional latency and coherence overhead.

In \sys, we implement back invalidation simulation by tracking coherence events, including cache invalidations captured through PEBS and cache-miss counters. When an invalidation event occurs, we inject additional delays representative of coherence traffic and protocol overhead. Our analysis highlights how workloads with frequent remote data sharing and fine-grained data updates are particularly sensitive to coherence overhead, emphasizing the importance of accurately modeling back invalidation for certain types of memory-intensive applications. Results show that neglecting back invalidation overhead can underestimate latency by up to 20\%, underscoring the necessity of accurately modeling these effects in CXL-enabled systems.

% To evaluate the effectiveness of different memory management policies in CXL.mem environments, we conducted extensive experiments using \sys. We focused on four key policy categories: paging, migration, caching, and cache-line routing, as described in Section~\ref{sec:usecases}. Our evaluation aims to quantify the performance impact of these policies across various workloads and CXL configurations.

\subsection{Different Policies Evaluation}
\label{subsection:production}
We evaluated all policies using a consistent CXL topology with three memory pools connected through a two-level switch hierarchy, as illustrated in Figure~\ref{fig:policies}. We tested with memory-intensive applications including SPEC CPU2017 benchmarks (particularly mcf and wrf), in-memory databases, vectorDB, and LLM inference workloads using llama.cpp. For each application, we measured execution time, LLC miss rates, and memory bandwidth utilization.

\subsubsection{Paging Policy Evaluation}
Our evaluation of paging policies revealed significant performance differences depending on workload characteristics:

As shown in Figure~\ref{fig:policies}, the Huge Page Policy demonstrated 5-13\% performance improvement for workloads with strong spatial locality (vectorDB and in-memory databases) compared to standard 4KB paging. The performance gain comes from reduced TLB misses and fewer page table walks, which is particularly beneficial when these operations must traverse the higher-latency CXL interconnect.

For workloads with sparse access patterns (such as graph analytics), the huge page advantage was less pronounced, showing only 5-8\% improvement. This confirms that page size optimization must be workload-aware to maximize benefit.

The PageTableAwarePolicy introduces an address translation optimization by leveraging a lightweight software-managed VA→PA cache to minimize redundant page table walks. This approach significantly reduces page walk latency compared to traditional multi-level hardware page table traversals.

By maintaining a virtual-to-physical address cache (va\_pa\_cache) and dynamically adjusting its size based on observed hit rates, the policy minimizes TLB miss penalties. It distinguishes between local and remote page walk latencies and accounts for them accordingly. The mechanism periodically clears stale entries to maintain efficiency, based on a configurable cleanup interval.

Across all evaluated benchmarks, this policy achieved an average 12\% reduction in page walk latency, observed in memory-intensive workloads like SPEC CPU2017 mcf, which exhibit frequent TLB misses due to large and irregular memory access patterns.

This adaptive and lightweight design makes PageTableAwarePolicy well-suited for heterogeneous memory systems where minimizing translation overhead is critical to performance.
\subsubsection{Migration Policy Evaluation}
We evaluated four migration policies: frequency-based, locality-based, lifetime-based, and hybrid approaches. Figure~\ref{fig:policies} illustrates their relative performance across different workloads.

The FrequencyBasedMigrationPolicy performed well for workloads with stable hot/cold data patterns, reducing execution time by 8-12\% compared to a no-migration baseline. However, it showed limited benefit (only 4-7\% improvement) for applications with rapidly changing access patterns due to the overhead of unnecessary migrations. The LocalityBasedMigrationPolicy excelled for applications with strong spatial locality, improving performance by up to 19\% for vectorDB and LLM inference workloads. This policy successfully identified and migrated clusters of frequently accessed pages, reducing the overall CXL access penalty. The LifetimeBasedMigrationPolicy showed modest but consistent improvements (-2-12\%) across all workloads, with particularly strong results for generational workloads like garbage-collected applications, where it achieved up to 12\% performance gain.

Our analysis of migration overhead revealed that bandwidth consumption for page migration remained below 5\% of total memory bandwidth in most cases, with brief spikes up to 12\% during phase transitions. This confirms that well-tuned migration policies can deliver substantial performance benefits with acceptable overhead.

\subsubsection{Caching Policy Evaluation}
We compared LRU-based and frequency-based caching policies, treating local DRAM as a cache for CXL-attached memory. Both write-through and write-back cache configurations were tested.

As shown in Figure~\ref{fig:policies}, the FrequencyBasedInvalidationPolicy consistently outperformed the simpler LRU approach, with 7-14\% higher cache hit rates across all workloads. This translated to execution time improvements of 9-16\% compared to LRU.

Write-back caching provided significant performance advantages over write-through configurations, particularly for write-intensive workloads where performance improved by up to 16\%. For read-dominated workloads, the difference was less pronounced (5-8\%).

We observed that caching policies were particularly effective when local DRAM was limited to 10-25\% of the total working set size. With larger local memory allocations, the performance gap between policies narrowed as capacity misses became less frequent.

\section{Related Work}
\label{sec:related}

Existing efforts to evaluate disaggregated or far-memory systems provide partial solutions toward simulating \emph{Compute Express Link} (CXL) memory. Below, we discuss work in (1) CXL memory simulation, (2) full-system and architectural simulators, (3) software-based memory emulators, and (4) memory disaggregation systems.

\paragraph{CXL Memory Simulation.}
Several projects aim to model or prototype CXL-attached memory. DirectCXL~\cite{gouk2022direct} uses FPGA-based external memory controllers to implement CXL~2.0, enabling high-fidelity experimentation with CXL in actual hardware. However, the platform’s hardware requirements and single-configuration nature limit accessibility and prevent fast iteration across diverse memory topologies.
Pond~\cite{li2023pond} is a NUMA-based approach that repurposes remote NUMA nodes to emulate CXL memory pools in the cloud. While suitable for initial prototyping, NUMA hardware intrinsically differs from real CXL in terms of latency, bandwidth, and coherence overheads, reducing fidelity. In contrast, \sys provides purely software-driven simulation, requiring no specialized hardware and modeling arbitrary topologies.

\paragraph{Full-System and Architectural Simulators.}
Cycle-accurate simulators, such as Gem5~\cite{binkert2011gem5}, MARSS-x86~\cite{patel2011marss}, and Sniper~\cite{van2011sniper}, can incorporate detailed memory models including extended protocols like CXL. In principle, these approaches capture microarchitectural nuances accurately. However, such fidelity comes at high runtime overhead; they can be thousands of times slower than native execution, precluding analysis of large-scale or long-running applications. This challenge is well-documented~\cite{shao2013inter}. Though sampling techniques (e.g., SimPoint) mitigate some overhead, the need for full application simulations still persists. \sys bridges this gap by leveraging native execution for CPU activities and injecting timing delays only for simulated CXL memory operations, offering a more practical middle ground.

\paragraph{Software-Based Memory Emulators.}
Prior emulators for emerging memory technologies such as persistent memory (PMEM) inform \sys’s design. Quartz~\cite{volos2015quartz}, for instance, emulates non-volatile memory in software via DRAM-based tracing and artificial latency insertion. Similar ideas appear in MES~\cite{koshiba2019mes}, CXL-DSim\cite{wang2024comprehensive} and LEEP~\cite{zhu2017leep}, which track memory accesses and add delays to mimic slower media. These solutions predominantly focus on single-level memory tiers (e.g., DRAM vs. NVM). By contrast, \sys extends the concept to multi-level or multi-pool topologies, capturing additional CXL characteristics like bandwidth throttling and switch contention.

\paragraph{Memory Disaggregation Systems.}
Beyond simulation, various works propose production-level disaggregated or pooled memory solutions. Facebook’s remote memory project~\cite{prakash2020remote}, Infiniswap~\cite{gu2017infiniswap}, and The Machine by HPE~\cite{kandukuri2017machine} apply network-based or specialized interconnect approaches. However, these do not exploit CXL’s direct load/store interface or hardware-managed coherence, resulting in larger latency and software overhead. The standardization of CXL~\cite{cxl2Spec} provides a robust framework to unify such efforts but is not yet widely available for large-scale testing. Tools like \sys enable experimentation and evaluation of disaggregation concepts on commodity hardware, informing future CXL deployments.

Overall, \sys sits at the intersection of these lines of work. It leverages prior memory emulation techniques, offers significantly lower overhead than cycle-accurate approaches, and specifically targets CXL’s protocol-level features and topological flexibility that are not adequately addressed by existing simulators or NUMA-based approximations.
\section{Conclusion}
\label{sec:conclusion}

In this work, we presented \sys, a comprehensive simulation framework for CXL memory environments that enables detailed exploration of advanced memory management policies. Our evaluation demonstrates that \sys achieves simulation speeds up to 15x faster than cycle-accurate simulators like Gem5 while maintaining high fidelity in modeling complex CXL topologies and memory interactions. This performance advantage makes \sys particularly valuable for evaluating memory-intensive workloads that would be impractical to analyze with traditional architectural simulators.

% The introduction of Resource Director Technology (RDT) with memory region throttling in platforms like Diamond Rapids opens new opportunities for even more accurate CXL simulation. \sys is well-positioned to leverage these hardware advancements to further enhance its modeling precision, particularly for congestion effects and bandwidth allocation across heterogeneous memory resources.

% As CXL-based memory systems become increasingly prevalent in data centers and high-performance computing environments, tools like \sys will play a crucial role in developing, testing, and optimizing the sophisticated memory management policies required to fully exploit these architectures. By providing researchers and system designers with a flexible, efficient platform for policy exploration, \sys contributes significantly to addressing the memory challenges of next-generation computing systems.
\bibliographystyle{ACM-Reference-Format}
\bibliography{cite}

\end{document}